\documentclass[a4paper,12pt]{article}
\usepackage{graphicx}
\usepackage{infnprep}
\usepackage[top=2.5cm, bottom=3cm, left=3cm, right=3cm]{geometry}

\newcommand{\Header}{
  \resizebox{15cm}{!}{
  \begin{tabular}{rl}
  \includegraphics[width=5cm, trim={50 100 0 0}]{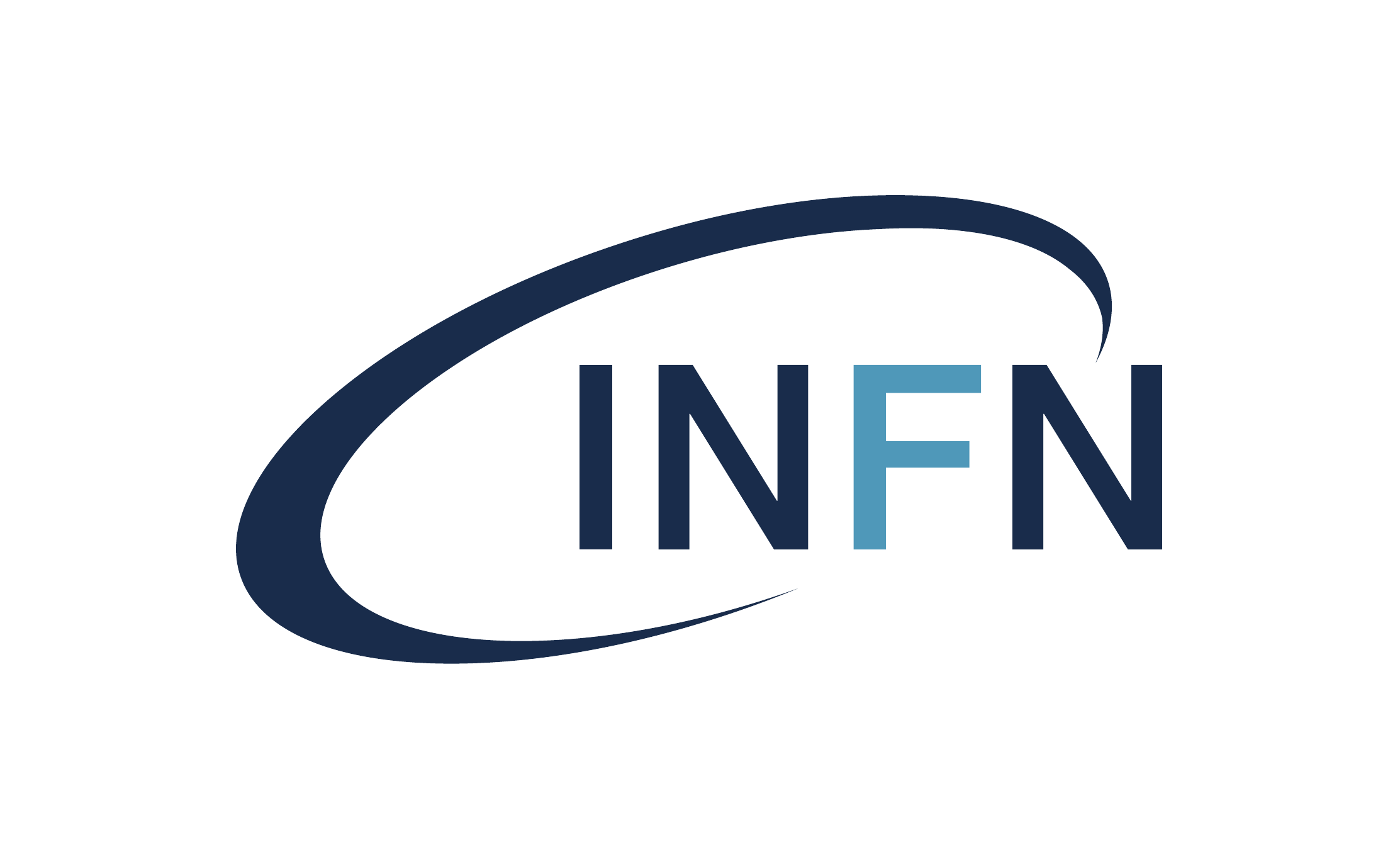} & {\LARGE\sffamily ISTITUTO NAZIONALE DI FISICA NUCLEARE}\\
      \\
  \end{tabular}
  }
    \renewcommand{\arraystretch}{1}
\vskip 0.5cm
\rule{15.0cm}{0.09mm}
\vskip 1.5cm
  \begin{flushright}
      {\underline{\bf INFN-19/1/LNF}}\\    
      {\small\bf 11 January 2019} \\      
  \end{flushright}
}

\begin{document}
\begin{titlepage}
\title
  {\Header \large \bf MPGD Optical Read Out for Directional Dark Matter Search}

\author{
G. Mazzitelli$^{1}$, E. Baracchini$^{1,2}$, G. Cavoto$^{2,3}$, E. Di Marco$^{2}$,\\
M. Marafini$^{2,4}$, C. Mancini$^{2}$,  D. Pinci$^{2}$, F. Renga$^{2}$ and S. Tomassini$^{1}$\\
{\it $^{1}$ INFN, Laboratori Nazionali di Frascati, Frascati (RM), Italy}\\
{\it $^{2}$ INFN, Sezione di Roma, Italy}\\
{\it $^{3}$ Sapienza Universit\`a di Roma, Dipartimento di Fisica, Italy}\\ 
{\it $^{4}$ Museo Storico della Fisica e Centro Studi e Ricerche E.Fermi, Italy}%
}

\maketitle

\baselineskip=14pt

\begin{abstract}
The Time Projection method is an ideal candidate to track low energy release particles. Large volumes can be readout by means of a moderate number of channels providing a complete 3D reconstruction of the charged tracks within the sensitive volume. It allows the measurement not only of the total released energy but also of the energy release density along the tracks that can be very useful for particle identification and to solve the head-tail ambiguity of the tracks. Moreover, gas represents a very interesting target to study Dark Matter interactions. In gas, nuclear recoils can travel enough to give rise to tracks long enough to be acquired and reconstructed.
\end{abstract}

\vspace*{\stretch{2}}
\begin{flushleft}
\end{flushleft}
\begin{flushright}
  \vskip 3cm
\small\it Published by \\
Laboratori Nazionali di Frascati\\
Conference Record of 2018 IEEE NSS/MIC/RTSD\\
presented by G.Mazzitelli
\end{flushright}
\end{titlepage}
\pagestyle{plain}
\setcounter{page}2
\baselineskip=17pt

\section{Introduction}
%
%
%
%
The use of large Time projection Chamber (TPC) in High Energy Physics (HEP) have various application and feature. Although with very small efficiency, those detectors offer among the best energy resolution and particle identification capability (PID), as well as a very good tracking and spacial resolution. This is pushing an international community, called CYGNUS, to study the application of such technology to the search of directional Dark Matter (DM) and the detection of neutrinos coming form the Sun.

In Italy the National Institute for Nuclear Physics (INFN) is promoting the Phase-0 for the construction of 1 m$^3$ demonstrator based on Micro Gas Pattern Detector (MGPD), namely a triple large Gas Electron Multiplier (GEM), optically read by means of a sCMOS low noise and high granularity sensor.

During the Phase-0 many prototypes, based on triple GEM optical readout, have been tested showing promising results to access the Phase-1 that will start in 2020 with the construction of the 1 m$^3$ demonstrator to be hosted in the National Laboratory of Gran Sasso (LNGS). A Phase-2 is foreseen to realize a 30-100 m$^3$ detector, as a brick of a world distributed observatory for DM and neutrinos within the CYGNUS international collaboration.  

Hence, The purpose of the Phase-0 is to demonstrate the validity and capability of large electron TCP at atmospheric pressure (He based) equipped with high ganularity and sensitivity MGPD optilal read out  

MGPD optical readout is relay changing the paradigm:
\begin{itemize}
\item optical sensors offer higher granularity with respect to electron sensitive devices;
\item optical coupling allows to keep sensor out of the sensitive volume reducing the interference with high
voltage operation and lowering the gas contamination;
\item the use of suitable lens allow to acquire large surfaces with small sensors;
\end{itemize}
Moreover, in last years, optical sensors had a huge development and are able to provide larger granularity along with very low noise level and high sensitivity.
Recently meany test has been done with different prototype (NITEC~\cite{JINST:nitec}, ORANGE~\cite{NIM:Marafinietal}, ~\cite{JINST:Marafinietal}, ~\cite{Antochi:2018otx},~\cite{Marafini:2016kzd}, LEMOn~\cite{Pinci:2017goi},~\cite{IEEE:Mazzitelli2017}, ~\cite{nim:pincielba}) at electron beam test facility, neutron beam and with various radioactive sources.

\section{Recent LEMOn Results}
\begin{figure}[!ht]
\centering
\includegraphics[width=5in]{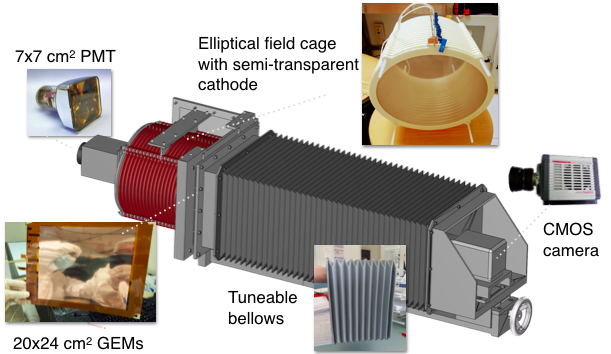}\DeclareGraphicsExtensions.
\caption{Engineering sketch of the tested prototype}
\label{fig:lemon_c}
\end{figure}

The LEMOn prototype is completely realize in 3D printing, (Fig.~\ref{fig:lemon_c}) and the heart consists of a 7 liter active drift volume surrounded by an elliptical field cage (200$\times$200$\times$240 mm$^3$) and a 200$\times$240 mm$^2$ rectangular triple GEM structure, with two 2 mm high transfer gaps among them. An ORCA-Flash 4.0 camera was placed 52 cm away from the last GEM to acquire images of light produced in GEM channels. This is based on a CMOS sensor with a high granularity (2048$\times$2048 pixels), very low noise (around two photons per pixel), high sensitivity (70\% of QE @ 600 nm) and linearity. This camera is instrumented with a Schneider lens (f/0.95-25 mm) and each pixel was looking at an effective area of 125$\times$125μm$^2$.

The prototype has been tested with the 450 MeV electron of the Frascati Beam Test facility~\cite{bib:btf} with a maximum drift field of 0.6 kV/cm and 60:40 HeCF$_4$ gas mixture:
\begin{itemize}
    \item an intrinsic XY resolution of about 70 $\mu$m, degraded of 10$\mu$m/cm as function of the drift distance
    \item an energy resolution  of 20-30\% in the keV energy range
    \item a drift diffusion  of about 130 $\mu$m/sqrt(cm) 
    \item a longitudinal (Z) resolution  exploiting the XY diffusion of 7\%
    \item a longitudinal (Z) resolution exploiting the PMT light diffusion of 10\%
\end{itemize}

\begin{figure}[!ht]
\centering
\includegraphics[width=5in]{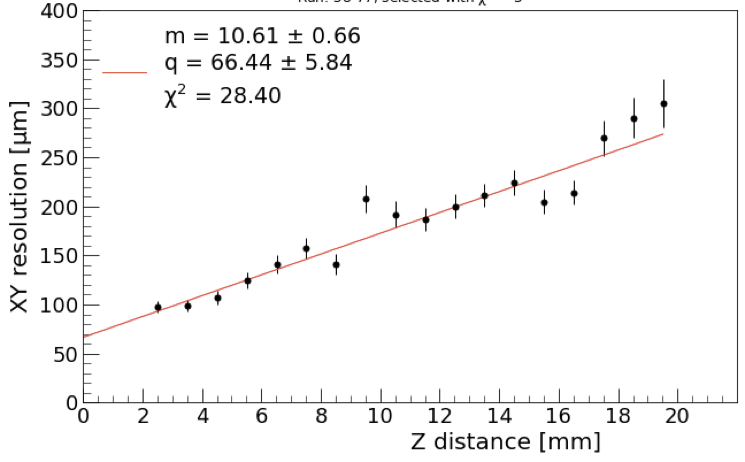}\DeclareGraphicsExtensions.
\caption{XY resolution vs depths (Z) obtained with a drift field to 0.6kV/cm}
\label{fig:XYres}
\end{figure}

The slop of 10$\mu$m/cm (Fig.~\ref{fig:XYres}) is strongly determined by the drift field and is limited in LEMOn by High Voltage connectors and the power supply. The new prototype, LIME, under construction at the National Laboratory of Frascati will be able to provide 1kV/cm over 50 cm of drift length. LIME will be also equipped with a 33$\times$33~cm GEM amplification stage constituting one of the 18 modules of the Phase-1 1 m$^3$ of CYGNO demonstrator.

\begin{figure}[!ht]
\centering
\includegraphics[width=5in]{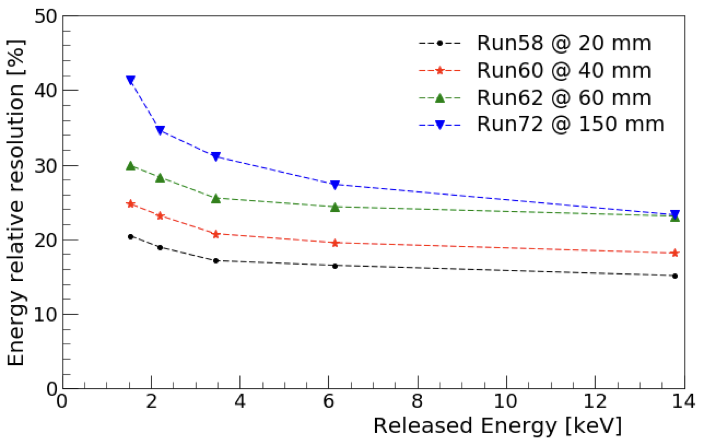}\DeclareGraphicsExtensions.
\caption{energy resolution @ depths (Z), in the few keV region a relative resolution of 20-30\% is achieved}
\label{fig:Eres}
\end{figure}

The energy resolution (Fig.~\ref{fig:Eres}) is obtained by cutting the 450 MeV single electron tracks in slides of given length and assuming an energy lost per sub-track length is 0.24 keV/mm as predicted by means of Garfield simulation for 60:40 HeCF$_4$ gas mixture.

\begin{figure}[!ht]
\centering
\includegraphics[width=5in]{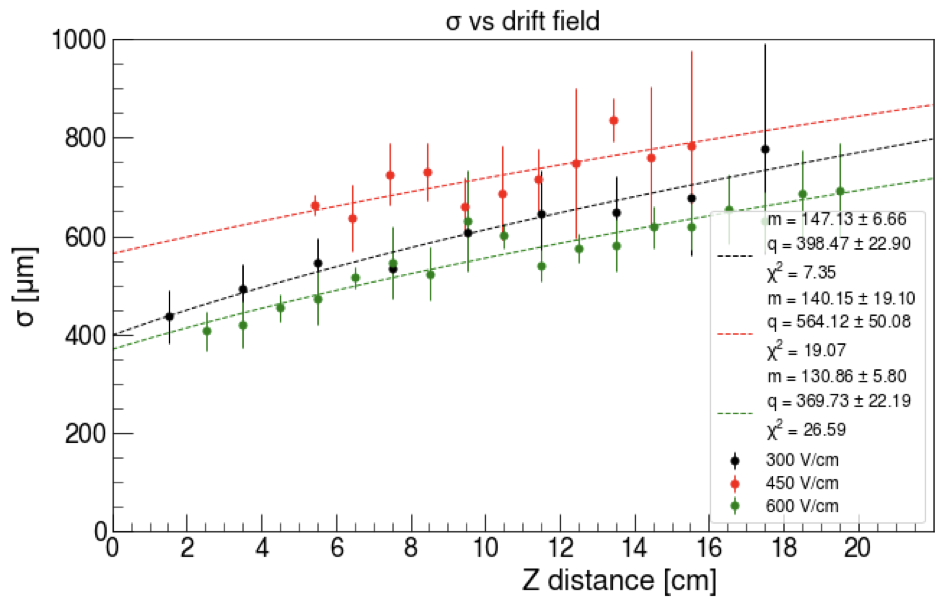}\DeclareGraphicsExtensions.
\caption{electrons drift diffusion in HeCF$_4$ for 0.30, 0.45, 0.60 kV/cm}
\label{fig:Diffusion}
\end{figure}

The diffusion (Fig.~\ref{fig:Diffusion}) is well in agreement with the expected value obtained with Garfiled simulation from where is also expected a better behaviour ad higher drift field up to 2kV/cm. In Fig.~\ref{fig:Diffusion}, data @ 0.45 kV/cm were collected with a different (not optimized) field applied on the triple GEM system, the intercept at zero of 400$\mu$m is typical for three GEM stack.

\begin{figure}[!ht]
\centering
\includegraphics[width=5in]{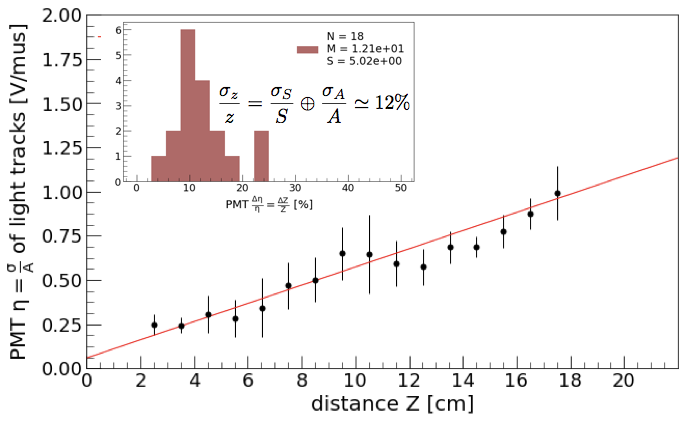}\DeclareGraphicsExtensions.
\caption{measured transverse diffusion and relative error $\sigma$Z $\simeq$ 12\%}
\label{fig:Ldiffusion}
\end{figure}

\begin{figure}[!ht]
\centering
\includegraphics[width=5in]{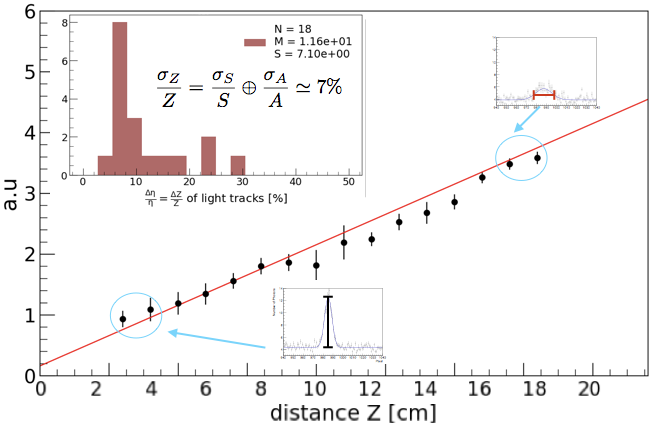}\DeclareGraphicsExtensions.
\caption{measured longitudinal diffusion and relative error $\sigma$Z $\simeq$ 7\%}
\label{fig:Tdiffusion}
\end{figure}

Transverse (Fig.~\ref{fig:Ldiffusion}) and longitudinal (Fig.~\ref{fig:Tdiffusion}) diffusion gives precision order 10\%. The exploitation of this characteristic could be very useful for the fiducialization of the detector: anode (where the GEM are located) and cathode are the most radioactive area in the detector, and a possible solution to avoid this noise is to trash the events close to them. A second method is to use minority carrier produced with negative ion by means of the introduction of SF$_6$ or CS$_2$ in the gas mixture. This solution, much more accurate and sophisticated see the support of an European Commission grant with the INITIUM project and will be investigated in Phase-1.

Moreover, a campaign of measurements with $^{55}$Fe has been started to study detection capability, energy resolution and efficiency at keV energy.

\begin{figure}[!ht]
\centering
\includegraphics[width=5in]{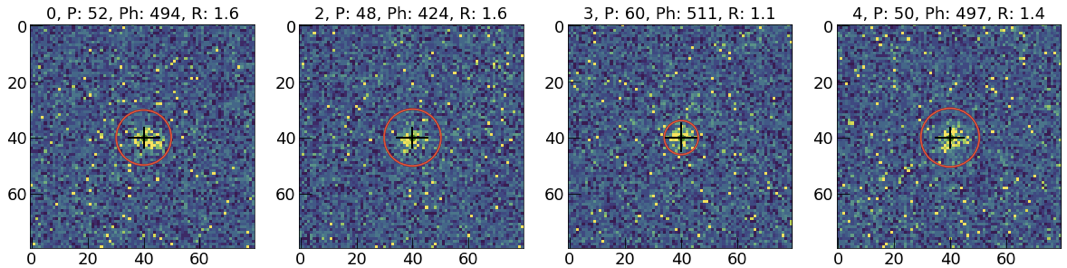}\DeclareGraphicsExtensions.
\caption{Example of $^{55}$Fe cluster identification obtained with NNC clustering methodology}
\label{fig:FEcluster}
\end{figure}

In order to investigate the detection capability different clustering methods are under study. The clustering method is fundamental to determinate the energy threshold and discriminate between $^{55}$Fe photons and background due to cosmic rays and environmental radioactivity that at see level for a 7 liters active volume start to be a significant noise. In the following the preliminary data obtained with the Nearest Neighbor Clustering (NNC) method are shown (Fig.~\ref{fig:FEcluster}).

The NCC method is CPU demanding and is not optimized at all on the possible shape of the track, is a zero order method used in parallel with HDBSCAN and Machine Learning approach that are under implementation to improve the energy threshold and PID.  

\begin{figure}[!ht]
\centering
\includegraphics[width=5in]{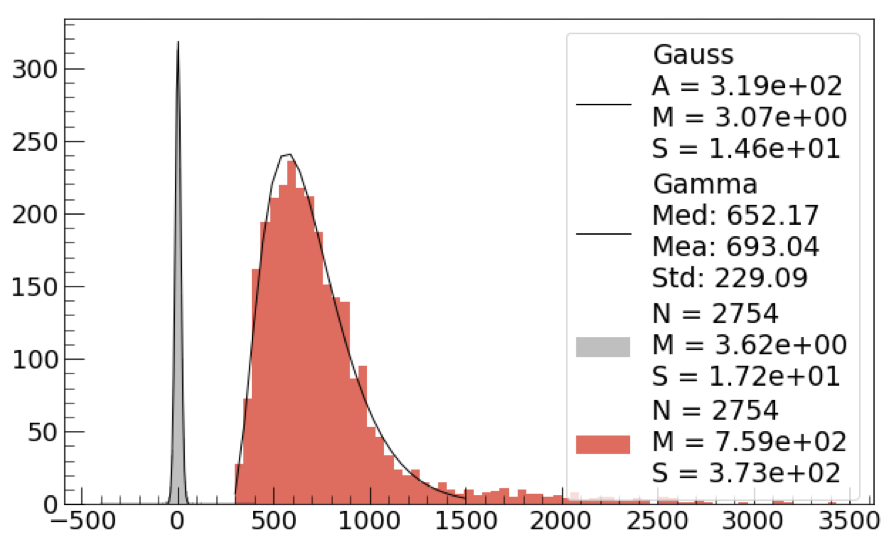}\DeclareGraphicsExtensions.
\caption{$^{55}$Fe energy spectrum obtained with NNC method obtained with GEM voltage of 460 V}
\label{fig:FEspectrum}
\end{figure}
Referring to Fig.~\ref{fig:FEspectrum}, the energy spectra fit provides 0.12 ph/eV. Assuming a pedestal jitters based on average of the run within $^{55}$Fe source in the detector of 5 $\sigma$ (75 photos), an energy threshold of 625 eV with a resolution of about 33\% (2 keV) in agreement with expected and measured value at the Frascati BTF. 
\section{Nuclear Recoil}
Last but not least, some preliminary tests have been performed with AmBe neutron source ad at the FNG ENEA neutron Facility~\cite{bib:fng}. A deep analysis of the data is ongoing, preliminary results show the very good capability of TPC-MGPD optical readout detector to identify particles.

\begin{figure}[!ht]
\centering

\includegraphics[width=5in]{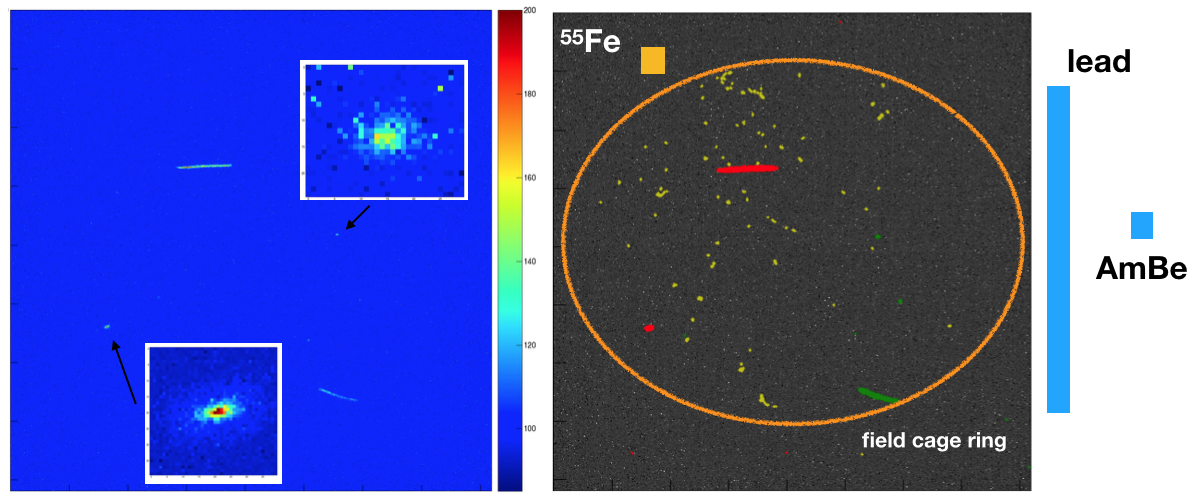}\DeclareGraphicsExtensions.
\caption{(left) example of row image collected with AmBe and $^{55}$Fe sources; (left/box) a zoom of millimeters high ionizing recoil (about 105 keV and 17 keV); (right) cluster reconstruction and Particle identification - ${55}$Fe Compton electrons and radioactivity background (yellow), p (green), $\alpha$ (red) - by means of average photons detected per pixel. The AmBe neutron source was located outside the detector on the right side of the picture; the orange elliptic rings represents the location of the field cage plastic rings}
\label{fig:ambe}
\end{figure}
In figure \ref{fig:ambe} a typical image collected with an AmBe neutron source is shown. The AmBe source was located at the right side of the image together with an $^{55}$Fe source located inside the detector in the opposite upper side. The AmBe source was shielded by means lead breaks in order to strongly reduce the associated emission of gamma and MeV electron and enhance the neutron interactions. In the right side of the figure the NCC cluster restriction is shown, where the different color, ${55}$Fe Compton electrons and radioactivity background (yellow), p (green), $\alpha$ (red), are identifying particles by means of average photons detected per pixel (less then 10 ph/pixel; between 10 and 15 ph/pixel; above 15 ph/pixel). The long nuclear recoil in the center of the figure (red) release about four time the energy of the particle (green) that candidate the first one to be an $\alpha$ and the second to be a proton produced by the interaction of neutron with the ASAP field cage ring, the yellow dots are due to $^{55}$Fe photon interactions, cosmic rays and radioactive background. In the box two zoom of a millimetric and nuclear recoil of about 100 keV and 20 KeV are shown.

\begin{figure}[!ht]
\centering

\includegraphics[width=5in]{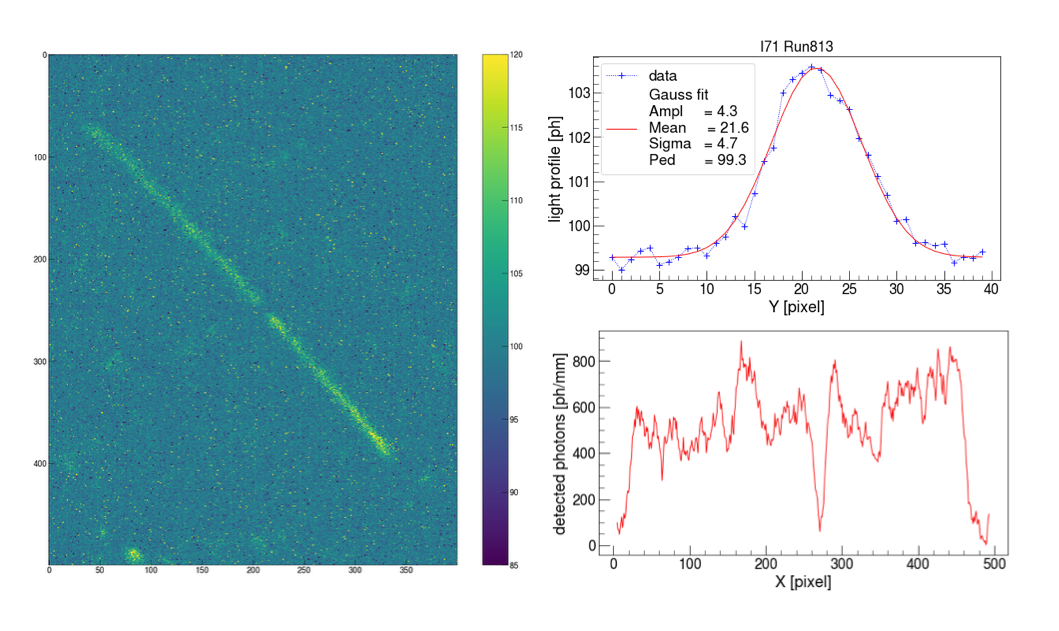}\DeclareGraphicsExtensions.
\caption{Nuclear Recoil in LEMOn (125 $\mu$m resolution) produced by a neutron of FNG ENEA source, head-tail and the brag pick are clearly visible}
\label{fig:proton}
\end{figure}

In figure \ref{fig:proton} a typical long nuclear recoil (namely a proton from the ASAP box) is shown where 550 ph/mm are released. Exploiting the results obtained with $^{55}$Fe calibration and GEM voltage of 440 V 0.06 ph/eV are expected, that means an energy deposition of 9.2 keV/mm. These results are consistent with the expectation of detecting highly ionizing millimeter traces with sufficient resolution, as required for a directional DM detector.

\section{Conclusion}
The work done during 2018 for the Phase-0 of the CYGNO project is confirming the potentiality of large TCP based equipped with an high granularity and sensitivity MPGD optical readout. Spacial end energy resolution, improvable with a stronger drift field, as well as the capability to identify nuclear recoil to avoid background looks very promising. Phase-0 is foreseen to be completed in 2019 with the constriction of LIME prototype, a module completely equivalent to 1/18 of the CYGNO 1 m$^3$ demonstrator.

\section*{Acknowledgment}
A special thanks to M. Chiti and A. Esposito of the Servizio Radioprotezione LNF-INFN and to A. Orlandi, M. Iannarelli and E. Paoletti of the Servizio progettazione rivelatori LNF-INFN. Morever, we are grateful to the BTF staff for the excellent and stable beam and the Frascati Enea FNG group, M. Pillon and S. Fiore, for their inexhaustible support.


\begin{thebibliography}{99}

\bibitem{JINST:nitec} 
    Baracchini, E. and Cavoto, G. and Mazzitelli, G. and
    Murtas, F. and Renga, F. and Tomassini, S.",
    Negative Ion Time Projection Chamber operation with
    SF$_6$ at nearly atmospheric pressure, JINST (2018), 13,P04022, doi: 10.1088/1748-0221/13/04/P04022
    
\bibitem{NIM:Marafinietal} 
    M.~Marafini {\it et al.}, 
    ``ORANGE: A high sensitivity particle tracker based on optically read out GEM,''
    Nucl.\ Instrum.\ Meth.\ A {\bf 845}, 285 (2017).
    doi:10.1016/j.nima.2016.04.014 

  \bibitem{JINST:Marafinietal}
    M. Marafini {\it et al.}, 
    ``High granularity tracker based on a Triple-GEM optically read by a CMOS-based camera,''
    JINST 10 (2015) 12, P12010 doi:10.1088/1748-0221/10/12/P12010.
    
  
\bibitem{Marafini:2016kzd} 
  M.~Marafini, V.~Patera, D.~Pinci, A.~Sarti, A.~Sciubba and E.~Spiriti,
  ``Optical readout of a triple-GEM detector by means of a CMOS sensor,''
  Nucl.\ Instrum.\ Meth.\ A {\bf 824}, 562 (2016).
  doi:10.1016/j.nima.2015.11.058
  
\bibitem{Pinci:2017goi}
  D.~Pinci {\it et al.},
  ``Cygnus: development of a high resolution TPC for rare events,''
  PoS EPS {\bf -HEP2017} (2017) 077.
  doi:10.22323/1.314.0077

\bibitem{Antochi:2018otx}
  V.~C.~Antochi {\it et al.},
  ``Combined readout of a triple-GEM detector,''
  JINST {\bf 13} (2018) no.05,  P05001
  doi:10.1088/1748-0221/13/05/P05001
  [arXiv:1803.06860 [physics.ins-det]].

\bibitem{IEEE:Mazzitelli2017}
    Mazzitelli, G. and E. Baracchini and M. Marafini and G.Cavoto and C. Voena and 
    F. Renga and E. Di Marco and D. Pinci and S. Tomassini and C. V. Antiochi,
    A high resolution TPC based on GEM optical readout, Under pubblication in IEEE Nuclear Science Symposuim Medical ImagingConference, 2017
\bibitem{nim:pincielba}
    D. Pinci, E. Baracchini, G. Cavoto, E. Di Marco, M. Marafini, G. Mazzitelli, F. Renga, S. Tomassini, C. Voena,
    High resolution TPC based on optically readout GEM,
    Nuclear Instruments and Methods in Physics Research Section A: Accelerators, Spectrometers, Detectors and Associated Equipment,
    2018, ISSN 0168-9002,https://doi.org/10.1016/j.nima.2018.11.085.
    
\bibitem{bib:btf}
    G.~Mazzitelli {\it et al}, 
    `` Commissioning of the DA$\Phi$NE beam test facility''
    Nucl. Instrum. Meth. A 515 (2003) 524–542.

 \bibitem{bib:fng}
     Martone, M., Angelone, M., Pillon, M., The 14 MeV Frascati neutron
     generator, Journal of Nuclear Materials 212-215(PART B), pp. 1661-1664 (1994)
\end{thebibliography}
\end{document}